\documentclass[psfig,showpacs,fleqn,nobibnotes]{revtex4}

\usepackage{amsmath}
\usepackage{graphicx}
\usepackage{float}
\usepackage{subfigure}

\newcommand{\rr}{\mbox{\boldmath $r$}}

\newcommand{\rb}{\mbox{\boldmath $b$}}

\begin{document}

\title{\bf Heavy  Quark Photoproduction in Coherent Interactions at High Energies}
\author{ V.P. Gon\c{c}alves $^{1}$, M.V.T. Machado  $^{2}$, A.R. Meneses $^{1}$ }
\affiliation{$^{1}$ \rm Instituto de F\'{\i}sica e Matem\'atica,  Universidade
Federal de Pelotas\\
Caixa Postal 354, CEP 96010-090, Pelotas, RS, Brazil\\
$^{2}$ \rm Centro de Ci\^encias Exatas e Tecnol\'ogicas, Universidade Federal do Pampa \\
Campus de Bag\'e, Rua Carlos Barbosa. CEP 96400-970. Bag\'e, RS, Brazil}

\begin{abstract}
 We
calculate the inclusive and diffractive photoproduction of heavy quarks in
proton-proton collisions at  Tevatron and LHC energies,  where the photon reaches energies larger than those ones  accessible at DESY-HERA.  The integrated cross section
and the rapidity distributions for charm and bottom production are computed within the color dipole picture employing three phenomenological saturation models based on the Color Glass Condensate formalism. Our results demonstrate that the experimental analyzes of these reactions is feasible and that the cross sections are sensitive to the underlying parton dynamics.
\end{abstract}

\maketitle

\section{Introduction}

The cross sections for heavy quark production in hadron-hadron and lepton-hadron collisions at high energies are strongly dependent on the behavior of the gluon distribution, which is determined by the underlying QCD dynamics (See e.g. Refs. \cite{reviewhq,mpla}). Theoretically, at high energies the QCD evolution leads to a system with high gluon density,  characterized by the limitation on the maximum phase-space parton density that can be reached in the hadron
wavefunction (parton saturation). The transition is specified  by a typical scale, which is energy dependent and is called saturation scale $Q_{\mathrm{sat}}$ (For recent reviews see Ref. \cite{cgc}).
Signals of parton saturation have already
been observed both in  $ep$ deep inelastic scattering at HERA and in deuteron-gold
collisions at RHIC (See, e.g. Ref. \cite{blaizot}).
In particular, in Ref. \cite{prl} we demonstrated that the inclusive charm total cross section  exhibits the property of geometric scaling , which is one of the main characteristics of the high density approaches.
However, the observation of this new regime still
needs confirmation and so there is an active search for new experimental signatures. In this paper  we study the inclusive and diffractive photoproduction of  heavy quarks in proton-proton collisions considering  three phenomenological models based on the Color Glass Condensate, which describe quite well the current experimental  HERA data for inclusive and exclusive observables. Our goal is twofold: update our previous studies \cite{vicmag_hq,vicmag_hqdif} considering these new parameterizations for the dipole scattering amplitude and present a  comparison  between the inclusive and diffractive production mechanisms using an identical theoretical input.

Our main motivation comes from the fact that in coherent interactions at Tevatron and LHC the photon reaches energies higher than those currently accessible at DESY - HERA. In hadron-hadron colliders, the relativistic protons  give rise to strong electromagnetic fields, which can interact with each other. Namely, quasi-real photons scatters off protons at very high energies in the current hadron colliders (For recent reviews on coherent interactions see, e.g., Ref. \cite{upc}). In particular, the heavy quark photoproduction cross section in a proton-proton collision is given by,
\begin{equation}
   \sigma(p+p \rightarrow p+ Q\overline{Q} + Y) = 2 \int_{0}^{\infty} \frac{dN_{\gamma}(\omega)}{d\omega}
   \, \sigma_{\gamma p\rightarrow Q\overline{Q} Y}\left(W_{\gamma p}^2= 2\,\omega\,\sqrt{S_{NN}}  \right) \, d \omega \; ,
\label{eq:sigma_pp}
\end{equation}
where $\omega$ is the photon energy in the center-of-mass frame (c.m.s.), $W_{\gamma p}$ is the c.m.s. photon-proton energy and $\sqrt{S_{NN}}$ denotes the proton-proton c.m.s.energy.  The final state $Y$ can be  a hadronic state generated by the fragmentation of one of the colliding protons (inclusive production) or a  proton (diffractive production).  The  photon spectrum is given by  \cite{Dress},
\begin{eqnarray}
\frac{dN_{\gamma}(\omega)}{d\omega} =  \frac{\alpha_{\mathrm{em}}}{2 \pi\, \omega} \left[ 1 + \left(1 -
\frac{2\,\omega}{\sqrt{S_{NN}}}\right)^2 \right]
\left( \ln{\Omega} - \frac{11}{6} + \frac{3}{\Omega}  - \frac{3}{2 \,\Omega^2} + \frac{1}{3 \,\Omega^3} \right) \,,
\label{eq:photon_spectrum}
\end{eqnarray}
with the notation $\Omega = 1 + [\,(0.71 \,\mathrm{GeV}^2)/Q_{\mathrm{min}}^2\,]$ and $Q_{\mathrm{min}}^2= \omega^2/[\,\gamma_L^2 \,(1-2\,\omega /\sqrt{S_{NN}})\,] \approx (\omega/
\gamma_L)^2$, where  $\gamma_L$ is the Lorentz factor. The expression above is derived considering the Weizs\"{a}cker-Williams method of virtual photons and using an elastic proton form factor (For more detail see Refs. \cite{Kleinpp,Dress}). It is important to emphasize that the expression (\ref{eq:photon_spectrum})  is based on a heuristic approximation,
which leads to an overestimation of the cross section at high energies ( $\approx 11 \%$ at $\sqrt{s}=1.3$ TeV)  in comparison with the more rigorous derivation of the photon spectrum for elastic scattering on protons derived in Ref. \cite{kniehl}.   For a more detailed comparison among the different photon spectra see Ref. \cite{nys_fluxo}. As a photon stemming from the electromagnetic field
of one of the two colliding protons can interact with one photon of
the other proton (two-photon process) or can  interact directly with the other proton (photon-hadron
process), both possibilities has been studied in the literature. In principle, the experimental signature of these two processes is distinct and it can easily be separated. While in two-photon interactions we expect the presence of two rapidities gaps and no hadron breakup, in the inclusive heavy quark photon-hadron production the hadron target we expect only one rapidity gap and the dissociation of the hadron.  However, as shown in Ref. \cite{vicmag_hqdif}, the  diffractive heavy quark photoproduction, where we also expect the presence of two rapidity gaps in the final state, similarly to two-photon interactions, is an  important  background for two-photon interactions as well as for the  dedicated program to search evidence of the Higgs and/or new physics in central  double diffractive production processes \cite{martin}.

In what follows, we briefly review the description of the inclusive and diffractive photoproduction of heavy quarks within the dipole picture and the distinct phenomenological saturation models which will be used in our calculations (Section \ref{sec2}). Our results are compared with the DESY-HERA data \cite{h1data} and the predictions from the bCGC and IP-SAT models for this process are presented for the first time. In Section \ref{sec3} the numerical results for the rapidity $y$ of the produced states and their total cross sections are shown. A comparison on the order of magnitude of the cross sections for the distinct models is performed. Finally in Section \ref{sec4} we present our main conclusions.

\section{QCD dipole picture and Saturation Models}
\label{sec2}

The photon-hadron interaction at high energy (small $x$) is usually described in the infinite momentum frame  of the hadron in terms of the scattering of the photon off a sea quark, which is typically emitted  by the small-$x$ gluons in the proton. However, in order to describe inclusive and diffractive interactions and disentangle the small-$x$ dynamics of the hadron wavefunction, it is more adequate to consider the photon-hadron scattering in the dipole frame, in which most of the energy is
carried by the hadron, while the  photon  has
just enough energy to dissociate into a quark-antiquark pair
before the scattering. In this representation the probing
projectile fluctuates into a
quark-antiquark pair (a dipole) with transverse separation
$\rr$ long after the interaction, which then
scatters off the target \cite{dipole}. The main motivation to use this color dipole approach, is that it gives a simple unified picture of inclusive and diffractive processes. In particular,  in this approach  the inclusive heavy quark photoproduction cross section $[\gamma p \rightarrow Q\overline{Q}X]$ reads as,
\begin{eqnarray}
\sigma_{tot}\, (\gamma p \rightarrow Q\overline{Q}X) = 2\, \int d^2\rb \int d^2\rr
 \int dz \Psi_{\gamma}^*(\rr,z) \, \mathcal{N}(x,\rr,\rb)\, \Psi_{\gamma}(\rr,z)\,\,.
\label{totalcsinc}
\end{eqnarray}
Furthermore, the diffractive cross section for the process $\gamma p \rightarrow Q\overline{Q}p$ is given by
\begin{eqnarray}
\sigma_{tot}^D\, (\gamma p \rightarrow Q\overline{Q}p) = \int d^2\rb \,   \int d^2\rr
 \int dz \Psi_{\gamma}^*(\rr,z) \, \mathcal{N}^2(x,\rr,\rb)\, \Psi_{\gamma}(\rr,z)  \,\,.
\label{totalcsdif}
\end{eqnarray}
In the Eqs. (\ref{totalcsinc}) and (\ref{totalcsdif})  the
variable $\rr$ defines the relative transverse
separation of the pair (dipole),  $z$ $(1-z)$ is the
longitudinal momentum fractions of the quark (antiquark) and the function $\Psi_{\gamma}(\rr,z)$ is the light-cone wavefunction for  transversely polarized photons, which depends in  our case of the  charge ($e_Q$) and mass ($m_Q$) of the heavy quark and is given by
\begin{eqnarray}
 |\Psi_{\gamma} (\rr,z)|^2 & = &\!  \frac{6\alpha_{\mathrm{em}}}{4\,\pi^2} \,
  e_Q^2 \, \left\{[z^2 + (1-z)^2]\, m_Q^2 \,K_1^2(m_Q \,\rr)
 +\,  m_Q^2 \, \,K_0^2(m_Q\,\rr) \,\,.
 \right\}  \,\,.
\label{wtrans}
 \end{eqnarray}

The  function $ {\cal N} (x, \rr, \rb)$ is the forward dipole-target scattering amplitude for a dipole with
size $\rr$ and impact parameter $\rb$ which encodes all the
information about the hadronic scattering, and thus about the
non-linear and quantum effects in the hadron wave function (see e.g. \cite{cgc}). It  can be obtained by
solving the BK (JIMWLK) evolution equation in the rapidity $Y \equiv \ln (1/x)$.
Many groups have studied the numerical solution of the BK equation, but
several improvements are still necessary  before using the solution in the calculation
of  observables. In particular, one needs to include the next-to-leading order
corrections into the evolution equation and perform a global analysis of all
small $x$ data. It is a program in progress (for recent results see \cite{alba,alba2}).
In the meantime it is necessary to use phenomenological models for
$ {\cal N}$ which capture the most essential properties of the solution.

During  the last years an intense activity in the area resulted  in
sophisticated models for the dipole-proton scattering amplitude, which have strong  theoretical
constraints and which are able to  describe the
HERA and/or RHIC data \cite{GBW,bgbk,iim,kowtea,kmw,pesc,watt08,kkt,dhj,gkmn,buw}. In what follows we will use
three distinct phenomenological saturation models based on the Color Glass Condensate which describe quite well the more recent HERA data: the IIM \cite{iim}, the bCGC \cite{kmw,watt08} and the IP-SAT model \cite{kowtea,kmw,watt08}.
In the IIM model \cite{iim} the scattering amplitude
${\mathcal N} (x,\rr,\rb)$ was constructed to smoothly interpol between the  limiting behaviors analytically under control: the solution of the BFKL equation
for small dipole sizes, $\rr\ll 1/Q_{\mathrm{sat}}(x)$, and the Levin-Tuchin law \cite{levin}
for larger ones, $\rr\gg 1/Q_{\mathrm{sat}}(x)$. Moreover, the authors have assumed that the impact parameter dependence can be factorized: ${\mathcal N} (x,\rr,\rb) = {\mathcal N} (x,\rr) S(\rb)$. A fit to the structure function $F_2(x,Q^2)$ was performed in the kinematical range of interest, showing that it is  not very sensitive to the details of the interpolation (For details see e.g \cite{vicmag_hq}). The predictions of this model for several observables were studied in Refs. \cite{forshaw_dif,fl_mag,fl_vicmag,vicmag_hq}.
Recently the IIM model was improved by the inclusion of  the impact parameter dependence in the scattering amplitude, with the resulting model being usually denoted bCGC.
 The parameters of this model were fitted to describe
the current HERA data in Ref. \cite{watt08}.  Following \cite{kmw} we have that the dipole-proton scattering amplitude  is given by:
\begin{eqnarray}
\mathcal{N}^{\mbox{bCGC}}(x,\rr,{\rb}) =   
\left\{ \begin{array}{ll} 
{\mathcal N}_0\, \left(\frac{ r \, Q_{s}}{2}\right)^{2\left(\gamma_s + 
\frac{\ln (2/r Q_{s})}{\kappa \,\lambda \,Y}\right)}  & \mbox{$r Q_{s} \le 2$} \\
 1 - \exp^{-A\,\ln^2\,(B \, r \, Q_{s})}   & \mbox{$r Q_{s}  > 2$} 
\end{array} \right.
\label{eq:bcgc}
\end{eqnarray}
with  $Y=\ln(1/x)$ and $\kappa = \chi''(\gamma_s)/\chi'(\gamma_s)$, where $\chi$ is the 
LO BFKL characteristic function.  The coefficients $A$ and $B$  
are determined uniquely from the condition that $\mathcal{N}(x,\rr)$  and its
derivative
with respect to $rQ_s$  are continuous at $rQ_s=2$. 
In this model, the proton saturation scale $Q_{s}$ now depends on the impact 
parameter:
\begin{equation} 
  Q_{s}\equiv Q_{s}(x,{\rb})=\left(\frac{x_0}{x}\right)^{\frac{\lambda}{2}}\;
\left[\exp\left(-\frac{{\rb}^2}{2B_{\rm CGC}}\right)\right]^{\frac{1}{2\gamma_s}}.
\label{newqs}
\end{equation}
The parameter $B_{\rm CGC}$  was  adjusted to give a good
description of the $t$-dependence of exclusive $J/\psi$ photoproduction.  
Moreover the factors $\mathcal{N}_0$ and  $\gamma_s$  were  taken  to be free. In this 
way a very good description of  $F_2$ data was obtained. 
The parameter set  which is going to be used here is the one presented in the second 
line of Table II of \cite{watt08}:  $\gamma_s = 0.46$, $B_{CGC} = 7.5$ GeV$^{-2}$,
$\mathcal{N}_0 = 0.558$, $x_0 = 1.84 \times 10^{-6}$ and $\lambda = 0.119$.
Furthermore, we will use in our calculations the scattering amplitude scattering proposed in Ref. \cite{kmw}, denoted IP-SAT,  which is given by
\begin{eqnarray}
\mathcal{N}^{\mbox{IP-SAT}}(x,\rr,{\rb}) = \left[1 - \exp\left( - \frac{\pi^2}{2\,N_c} \rr^2 \alpha_s(\mu^2)\,xg(x,\mu^2) \,T(\rb)\right)\right] \,\,,
\label{nkmw}
\end{eqnarray}
where the scale $\mu^2$ is related to the dipole size $\rr$ by $\mu^2 = 4/\rr^2 + \mu_0^2$ and the gluon density is evolved from a scale $\mu_0^2$ up to $\mu^2$ using LO DGLAP evolution without quarks  assuming that the initial gluon density is given by $xg(x,\mu_0^2) = A_g \, x^{-\lambda_g}\,(1 - x)^{5.6}$. The values of the parameters $\mu_0^2$, $A_g$ and $\lambda_g$ are determined from a fit to $F_2$ data. Moreover, it is assumed that the proton shape function $T(\rb)$  has a Gaussian form, $T(\rb) = 1/(2\pi B_G)\exp[-(\rb^2/2B_G)]$, with $B_G$ being a free parameter which is fixed by the fit to the differential cross sections for exclusive vector meson production. The parameter set used in our calculations is the one presented in the first line of Table III of \cite{kmw}:  $\mu_0^2 = 1.17$ GeV$^2$, $A_g = 2.55$,  $\lambda_g = 0.020$ and $B_G = 4$ GeV$^{-2}$.

As discussed in Refs. \cite{raju_ea4,raju_ea5} the expression (\ref{nkmw}) for the forward scattering amplitude can be obtained to leading logarithmic accuracy in the classical effective theory of the Color Class Condensate formalism. Moreover, it is applicable when the leading logarithms in $Q^2$ dominate the leading logarithms in $1/x$, with the small $\rr$ limit being described by the linear DGLAP evolution at small-$x$.  In contrast, the bCGC model for $\mathcal{N}$, Eq. \ref{eq:bcgc}, captures the basic properties of the quantum evolution in the CGC formalism, describing both the bremsstrahlung limit of linear small-$x$ evolution (BFKL equation) as well nonlinear renormalization group at high parton densities (very small-$x$). Consequently, the IP-SAT model can be considered a phenomenological model for the classical limit of the CGC, while the bCGC for the quantum limit. It is important to emphasize that both models provide excellent fits to a wide range of HERA data for $x\le 0.01$. Therefore, the study of  observables which are strongly dependent on $\mathcal{N}$ is very important to constrain the underlying QCD dynamics at high energies. In what follows we consider these two models as input in our calculations of the inclusive and diffractive heavy quark photoproduction in $\gamma p$ collisions at HERA and coherent $pp$ interactions at Tevatron and LHC energies. For comparison we also consider the IIM model \cite{iim}.

\begin{figure}
\begin{tabular}{cc}
\includegraphics[scale=0.5] {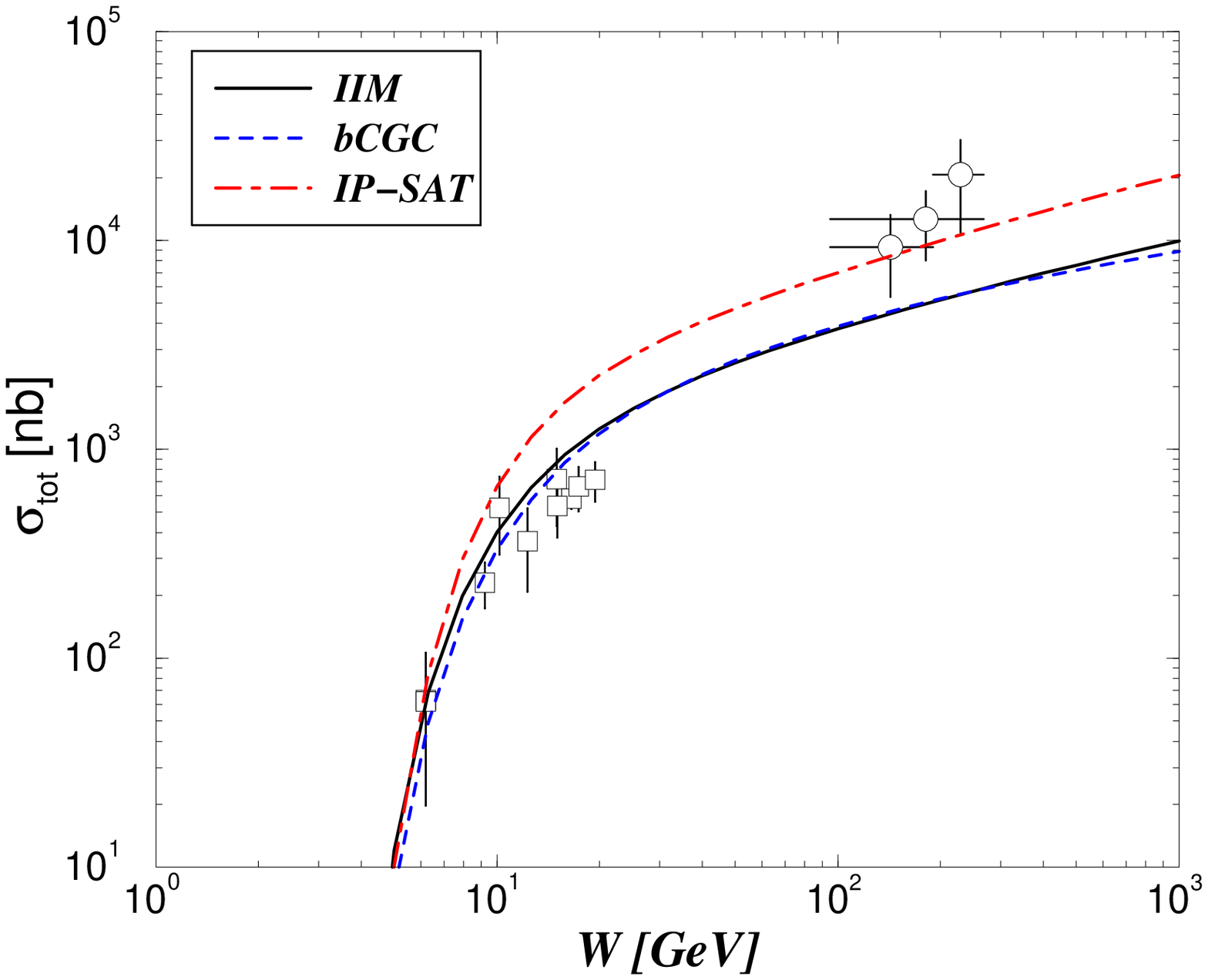} & \includegraphics[scale=0.5]{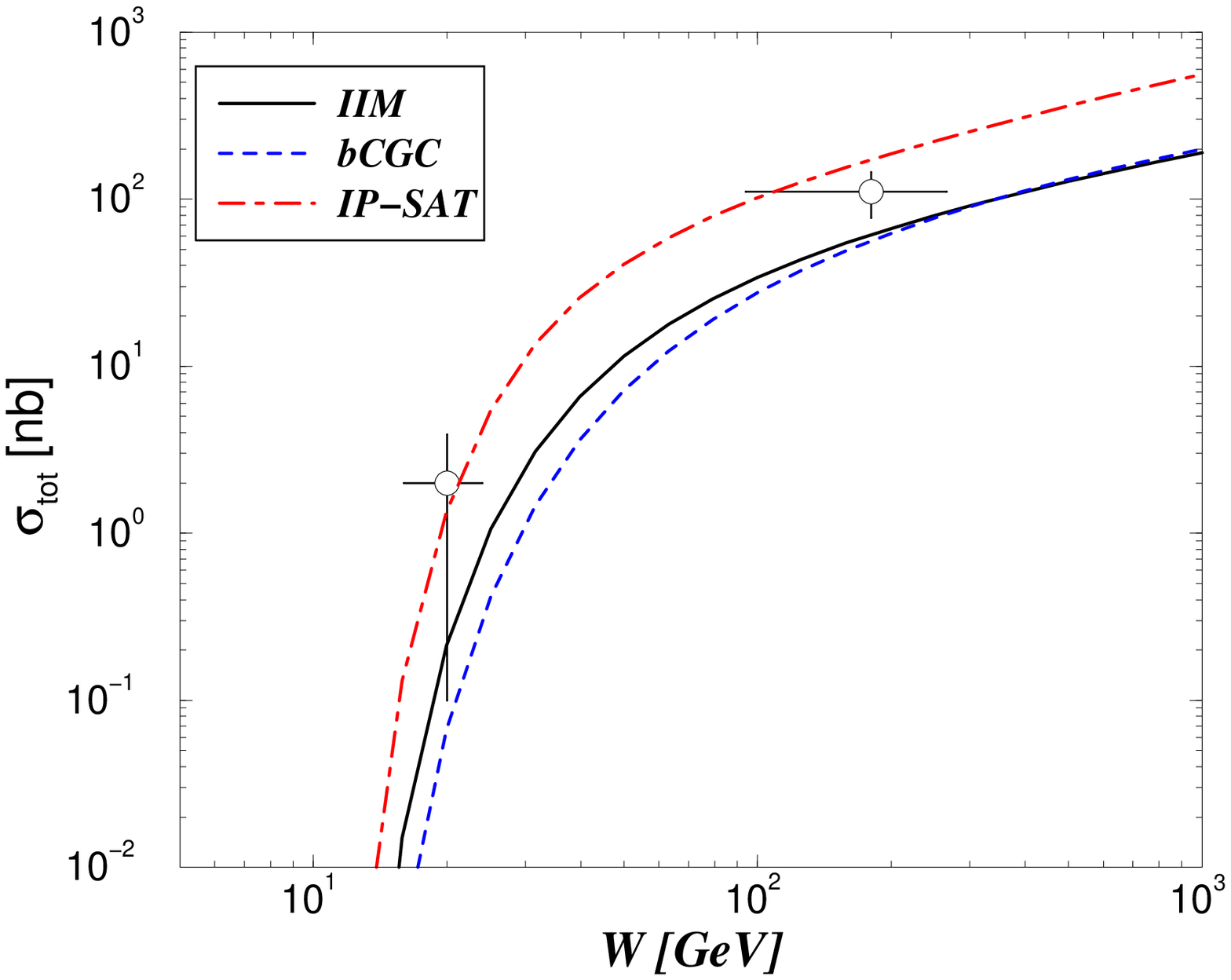}
\end{tabular}
\caption{ The total photoproduction cross section for charm (left panel) and bottom (right panel). The experimental measurements are from DESY-HERA. }
\label{fig1}
\end{figure}

 Having presented the phenomenological models which will be used in our calculations, in Fig. \ref{fig1} we compare the numerical results for the inclusive heavy quark photoproduction with the experimental DESY-HERA data \cite{h1data}. In all calculations we have use the same quark masses $m_c = 1.4$ GeV and $m_b = 4.5$ GeV. We quote Ref. \cite{vicmag_hq} for a comparison of the experimental data with another theoretical approaches. In order to describe the threshold region, $W\rightarrow 2m_Q$, we have multiplied the cross sections by a factor $(1-x)^7$, following studies presented in Ref. \cite{ASK}. In charm case (left panel), the IIM and bCGC predictions are almost identical in all kinematical range. In contrast, the predictions for these two models for bottom production   differ at small energy. The IP-SAT predictions are approximately a factor 2 larger than the IIM and bCGC predictions. One have that the IIM and bCGC models underestimate the experimental data for charm production at high energies, producing a reasonable description of the region near threshold (low energies: $W \le 20$ GeV). In contrast, the IP-SAT model describes the high energy region but overestimate the low energy regime. For the bottom case, the value of $x$ which determines the magnitude of the saturation scale is not sufficiently small, which implies that the cross section is dominated by the linear regime of the scattering amplitude. The three models give a reasonable description of the scarce experimental data. Unfortunately, the current precision and statistics of the experimental measurements of the photoproduction cross section are either low to formulate  definitive conclusions about the robustness of the different saturation models presented here. More precise measurements could be pose stringent constraints on the energy dependence and overall normalization. Finally, it should be noticed that the present calculations concern only the direct photon contribution to the cross section, whereas the resolved component has been neglected. In some extent the results from the saturation models presented here let some room for this contribution. Details on its calculation and size of its contribution can be found, for instance,  in Ref. \cite{Timneanu_Motyka}.

\begin{figure}
\includegraphics[scale=0.5]{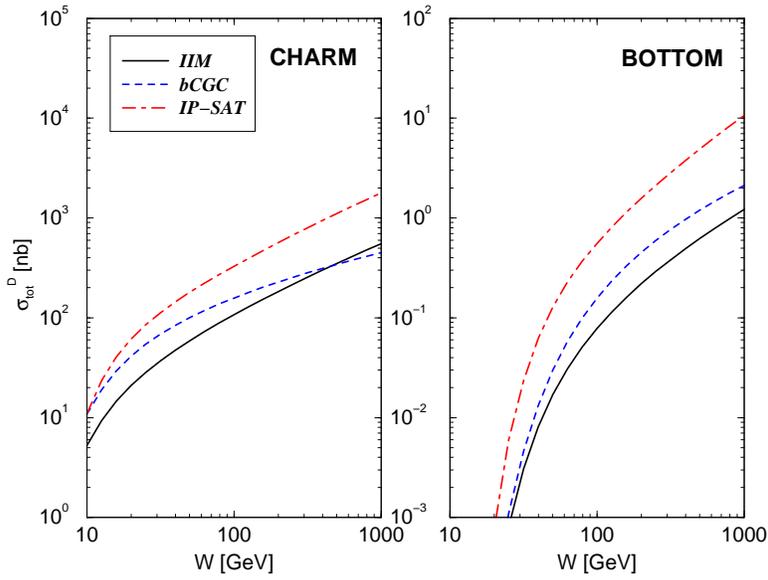}
\caption{ The energy dependence of the diffractive cross section for the charm (left panel) and bottom (right panel) production predicted by the distinct phenomenological models.}
\label{fig2}
\end{figure}

In Fig. \ref{fig2} we present our predictions for the diffractive photoproduction of heavy quarks. In comparison with the inclusive case, the  diffractive cross sections are approximately a factor 30 smaller. The main aspect is that the difference between the saturations models is enlarged, which is directly associated to the quadratic dependence of the cross section on the scattering amplitude. It implies that the experimental study of these observables can be useful to determine the QCD dynamics at high energies.

\section{Results}
\label{sec3}

In what follows, we will compute the rapidity distribution and total cross sections for the inclusive and diffractive photoproduction of heavy quarks  from proton-proton collisions at high energies. The phenomenological models shortly reviewed in the previous section serve as input for the numerical calculations using Eq. (\ref{eq:sigma_pp}) for  the energies of the  current and future $pp$ and $p\bar{p}$ accelerators.  Namely, one considers the  Tevatron value $\sqrt{S_{NN}}=1.96$ TeV for its $p\bar{p}$ running and for the planned LHC $pp$ running  one takes the design energy  $\sqrt{S_{NN}}=14$ TeV.

The distribution on rapidity $y$ of the produced open heavy quark state can be directly computed from Eq. (\ref{eq:sigma_pp}), by using its  relation with the photon energy $\omega$, i.e. $y\propto \ln \, (\omega/m_Q)$.  A reflection around  $y=0$ takes into account the interchanging between the proton's photon emitter and the proton target. Explicitly, the rapidity distribution is written down as,
\begin{eqnarray}
\frac{d\sigma \,\left[p+p \rightarrow p + Q\overline{Q} + Y) \right]}{dy} = \omega \, \frac{N_{\gamma} (\omega )}{d\omega }\,\sigma_{\gamma p \rightarrow Q\overline{Q}Y}\,\left(\omega \right)\,,
\end{eqnarray}
where $Y$ is a hadronic final state $X$ resulting of the proton fragmentation in the inclusive case and $Y = p$ for  diffractive production.

\begin{figure}
\begin{tabular}{cc}
\includegraphics[scale=0.3] {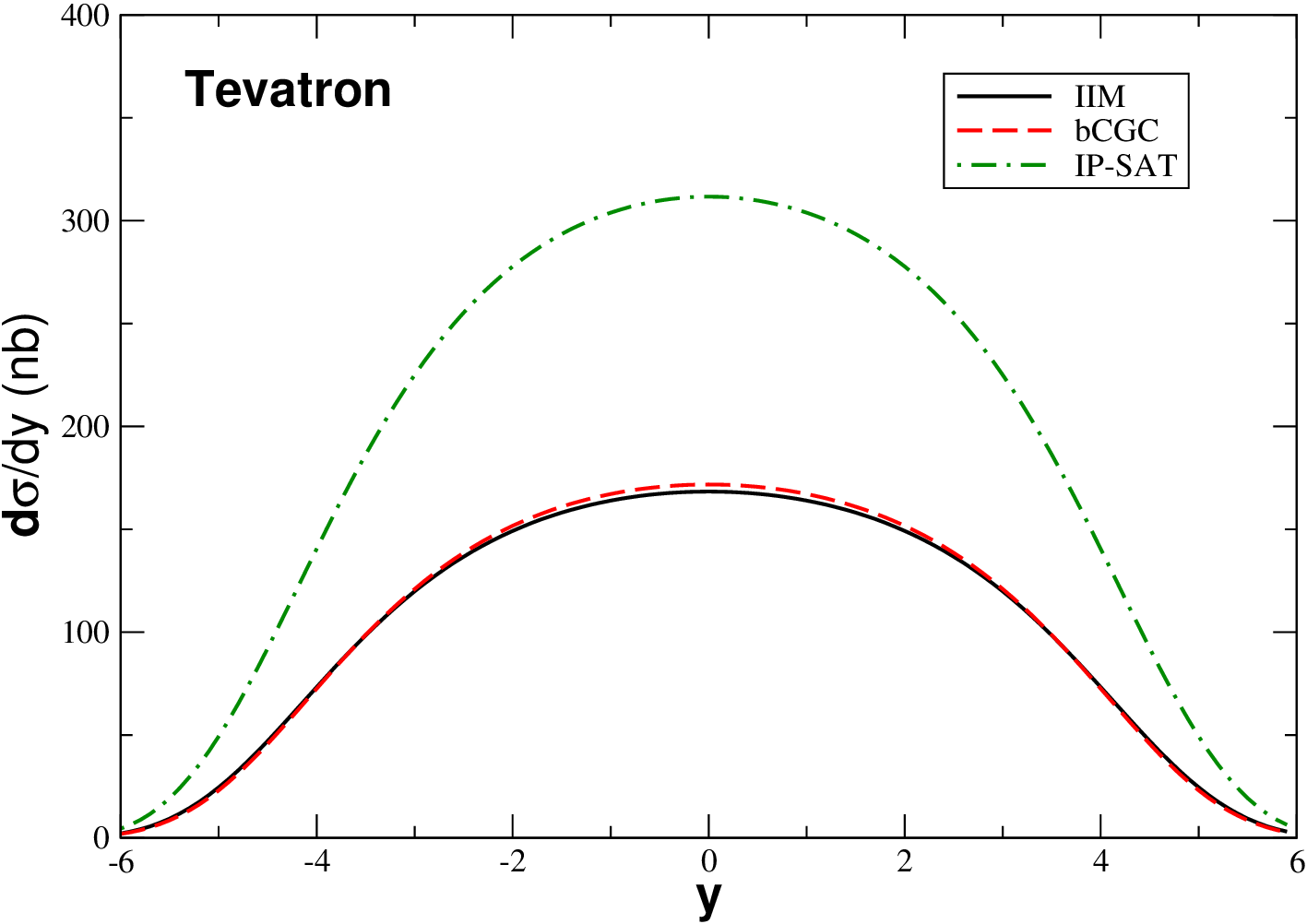} & \includegraphics[scale=0.3]{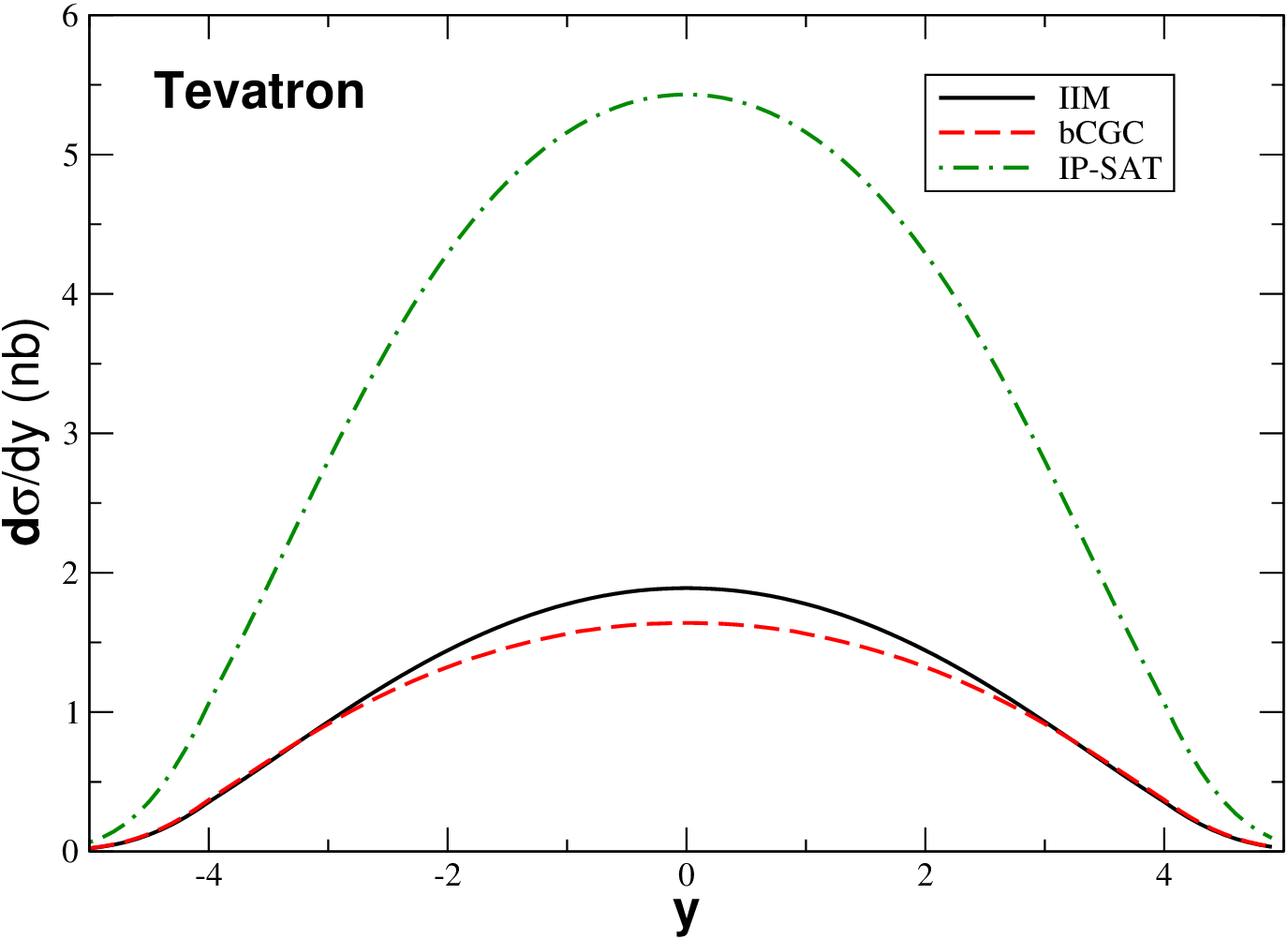}
\end{tabular}
\caption{ The rapidity distribution for the inclusive  charm (left panel) and bottom (right panel) photoproduction on $pp$ reactions at Tevatron energy $\sqrt{S_{NN}}=1.96\,\,\mathrm{TeV}$. Different curves correspond to distinct phenomenological models.}
\label{fig3}
\end{figure}

\begin{figure}
\begin{tabular}{cc}
\includegraphics[scale=0.3] {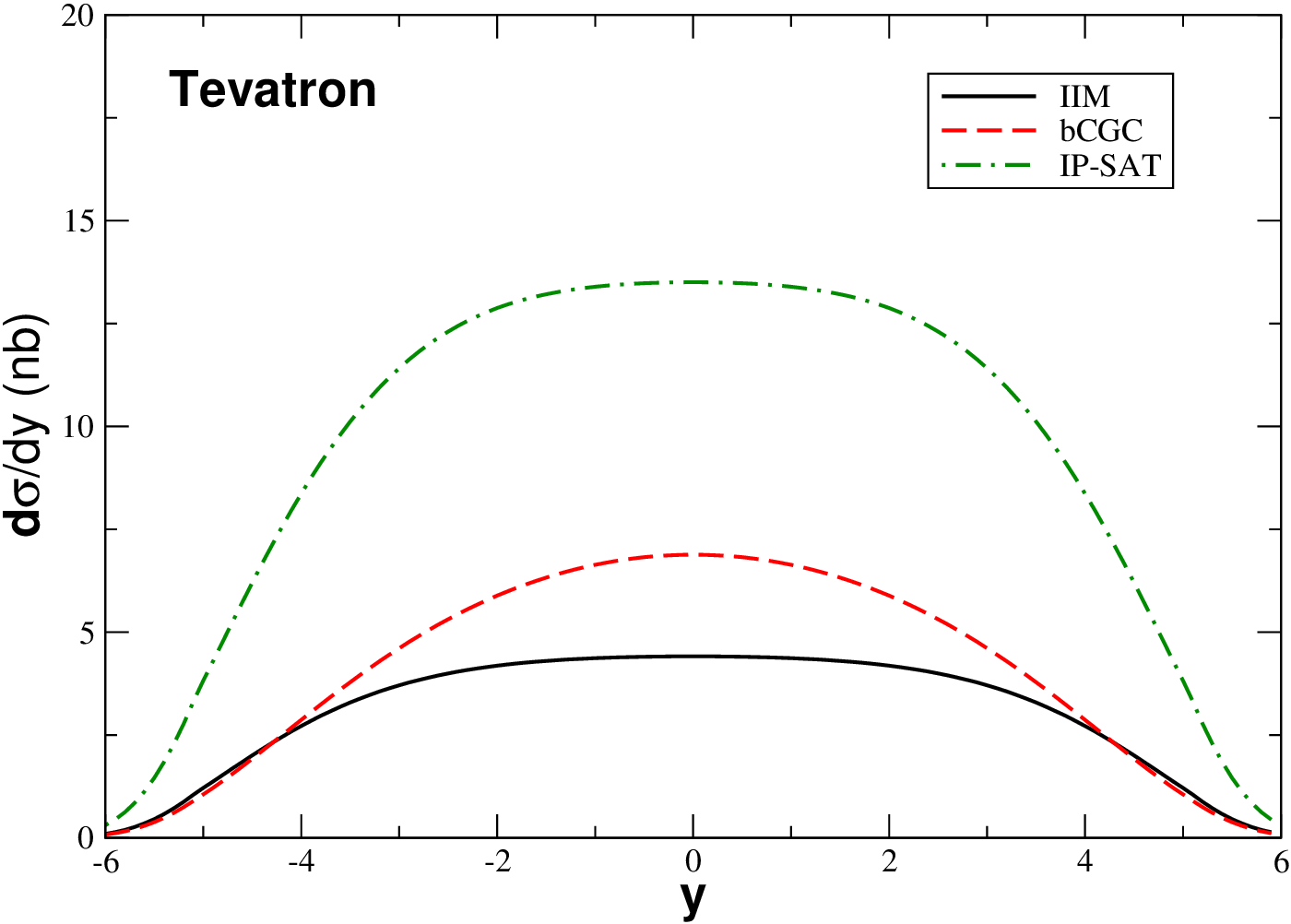} & \includegraphics[scale=0.3]{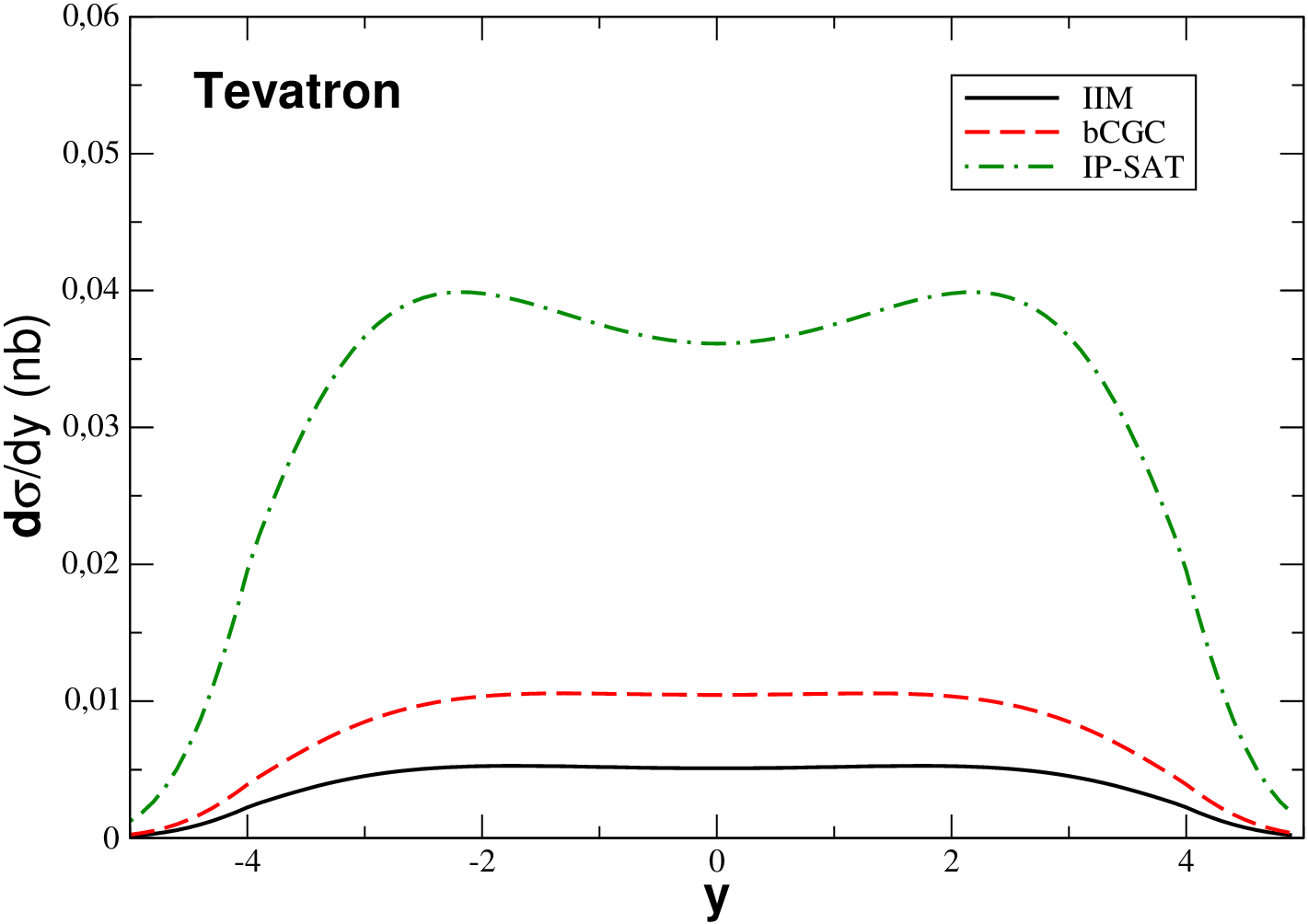}
\end{tabular}
\caption{ The rapidity distribution for the diffractive  charm (left panel) and bottom (right panel) photoproduction on $p\bar{p}$  reactions at Tevatron  energy $\sqrt{S_{NN}}=1.96\,\,\mathrm{TeV}$. Different curves correspond to distinct  phenomenological models.}
\label{fig4}
\end{figure}

\begin{figure}
\begin{tabular}{cc}
\includegraphics[scale=0.3] {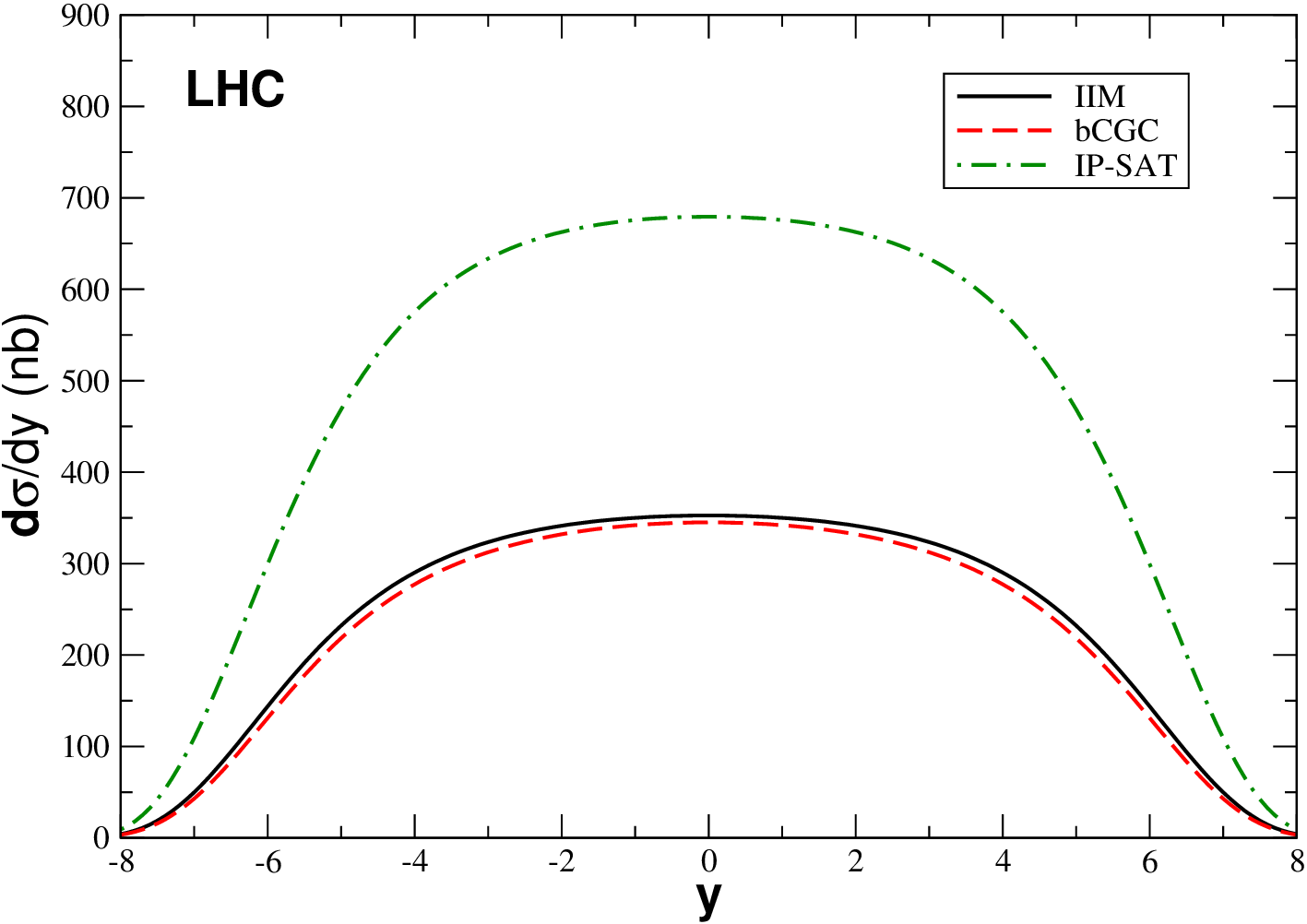} & \includegraphics[scale=0.3]{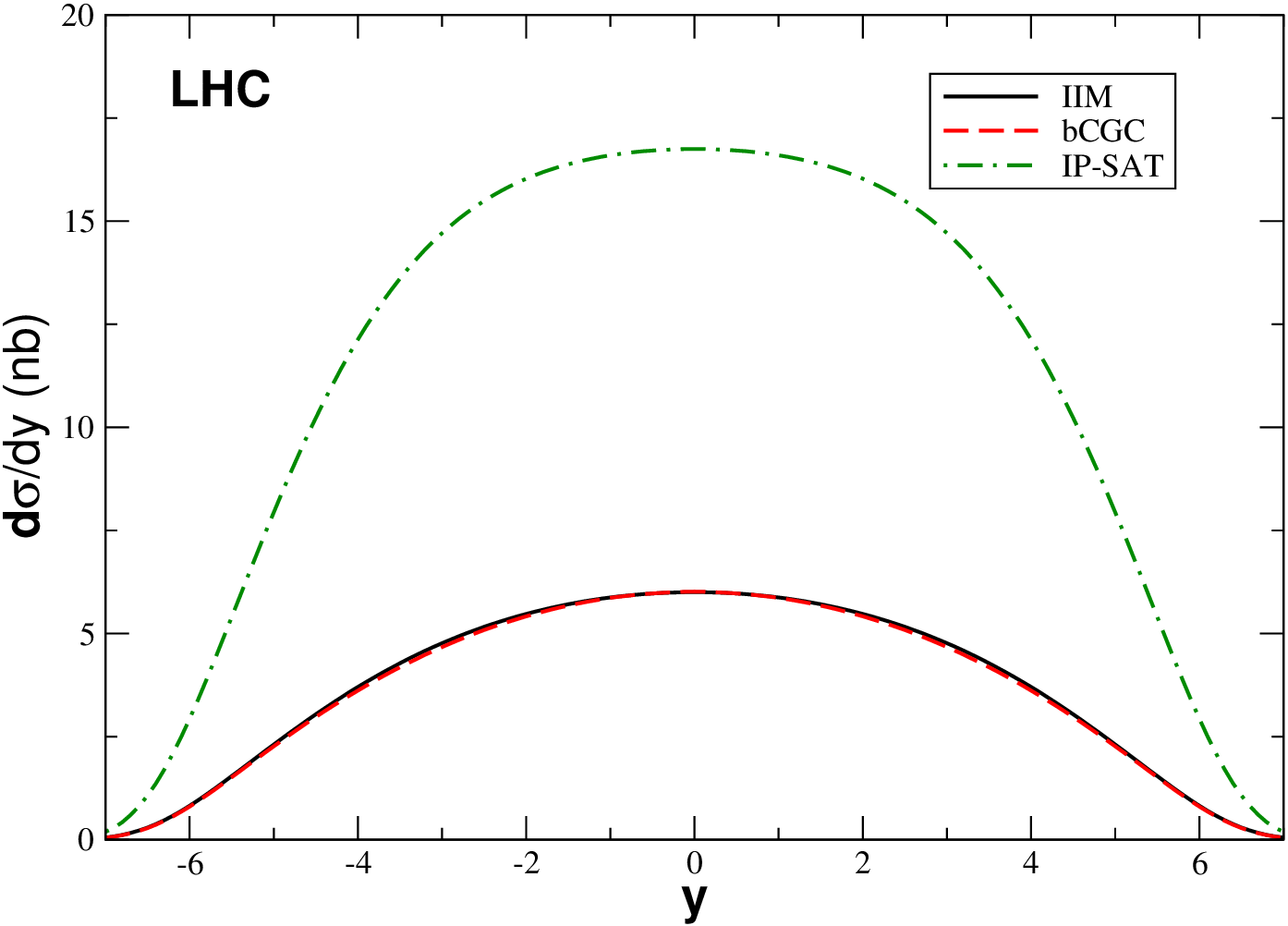}
\end{tabular}
\caption{ The rapidity distribution for the inclusive  charm (left panel) and bottom (right panel) photoproduction on $p\bar{p}$  reactions at LHC  energy $\sqrt{S_{NN}}=14\,\mathrm{TeV}$. Different curves correspond to distinct phenomenological models.}
\label{fig5}
\end{figure}

\begin{figure}
\begin{tabular}{cc}
\includegraphics[scale=0.3] {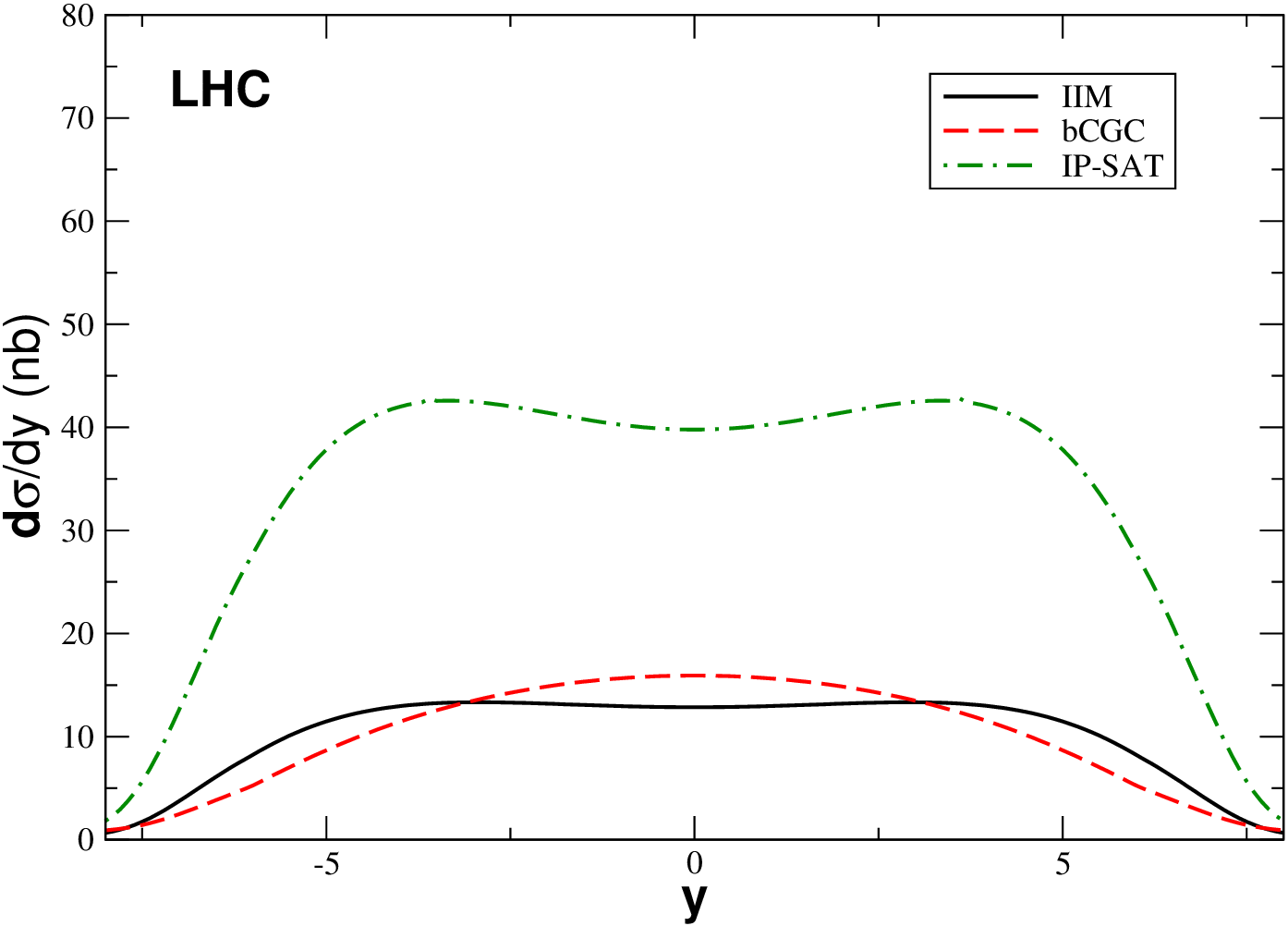} & \includegraphics[scale=0.3]{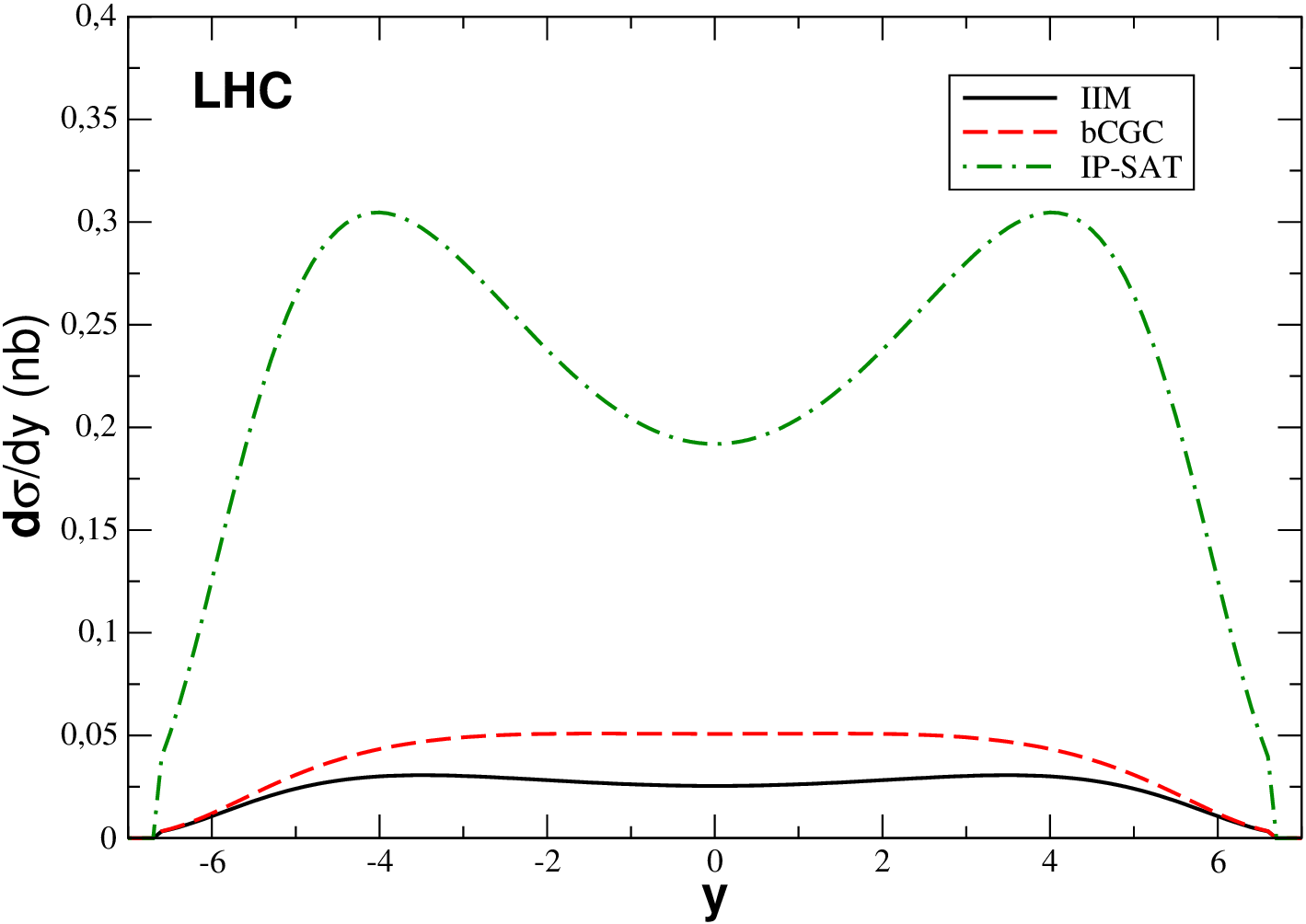}
\end{tabular}
\caption{ The rapidity distribution for the diffractive  charm (left panel) and bottom (right panel)  photoproduction on $p\bar{p}$  reactions at LHC   energy $\sqrt{S_{NN}}=14\,\mathrm{TeV}$. Different curves correspond to distinct phenomenological models.}
\label{fig6}
\end{figure}

The resulting rapidity distributions for inclusive and diffractive heavy quark photoproduction coming out of the distinct phenomenological models considered in previous section are depicted in Figs.  (\ref{fig3}-\ref{fig6}) at   Tevatron and LHC energies. For the inclusive case (Figs. \ref{fig3} and \ref{fig5}) the IIM and bCGC predictions are very similar, as expected from the analyzes at photon level in the previous section. In contrast, these predictions are distinct in the diffractive case, with the bCGC prediction being larger than IIM one at mid-rapidity. On the other hand, the IP-SAT prediction is larger than these predictions by a factor 2 (3) in the charm (bottom) case. We can consider the IIM and bCGC predictions as a lower bound for the coherent production of heavy quarks at Tevatron and LHC. Our results indicate that the experimental study of the inclusive heavy quark photoproduction can be very useful to discriminate between the classical and quantum versions of the CGC formalism. It also is true in the diffractive case, where the different models can be discriminated more easily.

\begin{table}[t]
\begin{center}
\begin{tabular} {||c|c|c|c|c||}
\hline
\hline
& $Q\overline{Q}$   & {\bf IIM} & {\bf bCGC} & {\bf IP-SAT}  \\
\hline
\hline
{\bf Tevatron} & $c\bar{c} \,\, (\mbox{incl.})$ &  1230 nb & 1245 nb & 2310 nb  \\
\hline
 & $c\bar{c} \,\, (\mbox{diff.})$ &  37 nb & 49 nb & 114 nb  \\
\hline
& $b\bar{b} \,\, (\mbox{incl.})$ &  11 nb & 10 nb & 32 nb  \\
\hline
 & $b\bar{b} \,\, (\mbox{diff.})$ &  0.04 nb & 0.08 nb & 0.30 nb  \\
\hline\hline
 {\bf LHC} & $c\bar{c} \,\, (\mbox{incl.})$ &  3821 nb & 3662 nb & 7542 nb  \\
\hline
 & $c\bar{c} \,\, (\mbox{diff.})$ &  165 nb & 161 nb & 532 nb  \\
\hline
& $b\bar{b} \,\, (\mbox{incl.})$ &  51 nb & 51 nb & 158 nb  \\
\hline
 & $b\bar{b} \,\, (\mbox{diff.})$ &  0.32 nb & 0.52 nb & 3 nb  \\
\hline
\hline
\end{tabular}
\end{center}
\caption{\ The integrated cross section for the inclusive and diffractive photoproduction of heavy quarks in $pp(\bar{p})$  collisions at Tevatron and LHC energies.}
\label{tabhq}
\end{table}

Let us now  compute the integrated  cross section considering the
distinct phenomenological models. The results are presented in Table \ref{tabhq},
for the inclusive and diffractive charm and bottom pair production at  Tevatron and LHC. The IP-SAT model gives
the  largest rates among the approaches studied, followed by the bCGC and IIM models with almost identical predictions, as a clear trend from the distribution on rapidity.
 In the inclusive case,  the values are either large at Tevatron and LHC, going from some units of nanobarns at Tevatron  to  microbarns at LHC. Therefore, these reactions can have high rates at the  LHC kinematical regime.
On the other hand, the cross sections for diffractive production  are approximately two orders of magnitude smaller than the inclusive case, but due the clear experimental signature of this process (two rapidity gaps), its experimental analyzes still is feasible.

In comparison with our previous results for the inclusive production of heavy quarks \cite{vicmag_hq} we have that  our predictions using the modern phenomenological IIM and bCGC models are similar. However, the IP-SAT prediction is factor of $\approx$ 2 larger. In the diffractive case, our predictions are larger by a factor $\gtrsim$ 2 than those presented in \cite{vicmag_hqdif}, where we have used the GBW model \cite{GBW} as input in our calculations. This behavior is directly associated to the different energy dependence predicted by the models for the linear regime. Furthermore, in comparison with the predictions for the heavy quark hadroproduction (See e.g. \cite{rauf}),  photoproduction cross sections are smaller ($\lesssim $ 1\%). However, the experimental separation between these two mechanics is feasible due to the presence of one rapidity gap in the photoproduction process.
 
 %

Lets now calculate the production rates for charm and bottom production in coherent interactions. At Tevatron, assuming the  design luminosity ${\cal L}_{\mathrm{Tevatron}} = 2\times \,10^{32}$ cm$^{-2}$s$^{-1}$, we have for inclusive production of charm $2-4 \times 10^{2}$ and for bottom $2-6$ events/second. In the diffractive case, we predict 7 - 22 ($8-60 \times 10^{-3}$)  events/second for charm (bottom) production. At  LHC, where ${\cal L}_{\mathrm{LHC}} = 10^{34}$ cm$^{-2}$s$^{-1}$), we predict for inclusive (diffractive)  charm  production $38-75 \times 10^{3}$ ($16-52 \times 10^{2}$) and for bottom $5 - 15 \times 10^{2}$ (3 - 30) events/second.
 Notice the large rate for bottom at LHC.

 Finally, lets discuss the experimental separation of the inclusive and diffractive photoproduction of heavy quarks. As emphasized in Refs. \cite{Kleinpp,vicmag_hq,vicmag_hqdif}, although the inclusive photoproduction cross section to be  a small fraction of the hadronic cross section, the separation of this channel is feasible if we impose the presence of a rapidity gap in the final state. It occurs due to the proton which is the photon emitter remains intact in the process. We expect that a cut in the transverse momentum of the pair could eliminate most part of the  contribution associated to the hadroproduction of heavy quarks.  Moreover, in comparison with the hadroproduction of heavy quarks, the event multiplicity for photoproduction interactions is lower, which implies that it may be used as a separation factor between these processes.
In the case of diffractive photoproduction of heavy quarks  
we expect the presence of two rapidity gaps in the final state, similarly to two-photon or Pomeron-Pomeron interactions. Consequently, it is important to determine the magnitude of this cross section in order to estimate the background for these other channels.  In particular, the central exclusive diffraction (CED) process characterized by the production of a final state  via fusion of two Pomerons has being intensely studied
as an alternative process to search evidence of the Higgs and/or new physics \cite{martin}, with the main background being the exclusive $b\overline{b}$ production. In Ref. \cite{mag_hqdif}, the double diffractive (DD)  heavy quark production is studied using the diffractive factorization theorem, including absorption corrections. The magnitude of the cross section is the following:  for Tevatron one has $\sigma_{c\bar{c}}^{\mathrm{DD}}\simeq 4.6$ $\mu$b and  $\sigma_{b\bar{b}}^{\mathrm{DD}}\simeq 0.1$ $\mu$b, whereas for the LHC one obtains $\sigma_{c\bar{c}}^{\mathrm{DD}}\simeq 18$ $\mu$b and $\sigma_{b\bar{b}}^{\mathrm{DD}}\simeq 0.5$ $\mu$b. It is expected that emerging protons from CED and DD processes have a much larger transverse momentum than those resulting from diffractive photoproduction processes. Consequently, in principle it is possible to introduce a selection criteria  to separate these two processes. However, this subject deserves more detailed studies.

\section{Conclusions}
\label{sec4}

In summary, we have computed the cross sections for inclusive and diffractive  photoproduction of heavy quarks in $p\bar{p}$ and $pp$  collisions at Tevatron and LHC energies, respectively. This has been performed using  modern phenomenological models based on the Color Glass Condensate formalism, which describe quite well the inclusive and exclusive observables measured in $ep$ collisions at HERA.
  The obtained values are shown to be sizeable  at Tevatron and are  increasingly larger at LHC. The feasibility of detection of these reactions is encouraging, since their experimental signature should be suitably clear.  Furthermore, they enable to constrain the underlying QCD dynamics at high energies, which is fundamental to predict the observables which will be measured in central hadron-hadron collisions at LHC.

\begin{acknowledgments}
 This work was partially financed by the Brazilian
funding agencies CNPq and FAPERGS.
\end{acknowledgments}

\newpage

\vspace{3cm}

\newpage

\vspace{3cm}

\newpage

\vspace{3cm}

\end{document}